\documentclass[aps,prd,twocolumn,superscriptaddress,nofootinbib]{revtex4-1}

\usepackage{amsmath,amssymb,amsfonts,bm,slashed,graphicx,longtable,array,multirow}
\usepackage{multirow, hyperref}
\usepackage{breakurl}
\usepackage{rotating}
\usepackage{soul}
\usepackage{comment}

\newcommand{\X}{\varphi}

\allowdisplaybreaks

\makeatletter
\g@addto@macro\bfseries{\boldmath}
\makeatother 

\graphicspath{{./figures/}}

\usepackage[dvipsnames]{xcolor}
     \definecolor{green}{rgb}{0,.3,0}
     \definecolor{red}{rgb}{.3,0,0}
     \definecolor{blue}{rgb}{0,0,.9}

\begin{document}

\title{Leveraging the ALICE/L3 cavern for long-lived exotics}

\author{Vladimir V. Gligorov}
\affiliation{LPNHE, Sorbonne Universit\'{e} et Universit\'{e} Paris Diderot, CNRS/IN2P3, Paris, France}

\author{Simon Knapen}
\affiliation{Ernest Orlando Lawrence Berkeley National Laboratory, University of California, Berkeley, CA 94720, USA}
\affiliation{Department of Physics, University of California, Berkeley, CA 94720, USA}
\affiliation{School of Natural Sciences, Institute for Advanced Study, Princeton, NJ 08540, USA}

\author{Benjamin Nachman}
\affiliation{Ernest Orlando Lawrence Berkeley National Laboratory, University of California, Berkeley, CA 94720, USA}

\author{Michele Papucci}
\affiliation{Ernest Orlando Lawrence Berkeley National Laboratory, University of California, Berkeley, CA 94720, USA}
\affiliation{Department of Physics, University of California, Berkeley, CA 94720, USA}
\affiliation{Theoretical Physics Department, CERN, Geneva, Switzerland}

\author{Dean J. Robinson}
\affiliation{Ernest Orlando Lawrence Berkeley National Laboratory, 
University of California, Berkeley, CA 94720, USA}
\affiliation{Santa Cruz Institute for Particle Physics and
Department of Physics, University of California at Santa Cruz,
Santa Cruz, CA 95064, USA}

\begin{abstract}
Run 5 of the HL-LHC era (and beyond) may provide new opportunities to search for physics beyond the standard model (BSM) at interaction point 2 (IP2). In particular, 
taking advantage of the existing ALICE detector and infrastructure provides an opportunity to search for displaced decays of beyond standard model long-lived particles (LLPs). 
While this proposal may well be preempted by ongoing ALICE physics goals, examination of its potential new physics reach
provides a compelling comparison with respect to other LLP proposals. In particular, full event reconstruction and particle identification could be possible by making use of the existing L3 magnet and ALICE time projection chamber. 
For several well-motivated portals, the reach competes with or exceeds the sensitivity of MATHUSLA and SHiP, provided that a total integrated luminosity of approximately $100\, \text{fb}^{-1}$ could be delivered to IP2.
\end{abstract}

\maketitle

\section{Introduction}

Interaction point 2 (IP2) at the Large Hadron Collider (LHC) accelerator complex is currently used by the \mbox{ALICE} experiment~\cite{Aamodt:2008zz} for the study of the quark-gluon plasma at high temperatures (examples of high temperature QCD discoveries achieved by ALICE can be found in e.g. Refs.~\cite{ALICE:2017jyt,Adam:2015pna}). The ALICE collaboration has firm plans to upgrade its detector and continue running throughout Run 3 and part of Run 4 \cite{Abelev:1475243}. However, should the heavy ion program conclude after Run 4 and with the long term future of the CERN accelerator program now taking shape, it would be remiss not to consider possible new opportunities at IP2 during Run 5 and beyond. 

The ALICE experiment comprises in part a gas Time Projection Chamber (TPC) detector housed within the L3 electromagnet~\cite{Wittgenstein1990}, and is designed to reconstruct very high multiplicities of tracks from ultra-relativistic ion-ion collisions. The L3 magnet has an interior cylindrical volume of length $12$\,m and radius $5.9$\,m, and a central field of $0.5$~T; the existing ALICE TPC~\cite{CERN-LHCC-2013-020} has radius 0.85\,m to 2.5\,m and a length of 5\,m along the beam axis.
The combination of a high resolution tracker -- the ALICE TPC and/or a larger one in this volume -- and the $0.5$\,T magnetic field would allow for both particle identification 
and momentum measurement, which would be tremendously advantageous for establishing an exotic particle discovery.  In this study we investigate the physics reach of a dedicated detector for the decay-in-flight of long-lived particles in this space. We refer to this hypothetical experiment as A Laboratory for Long-Lived eXotics (AL3X) (pronounced `Alex'). 

In many well-motivated theoretical frameworks, long lived particles (LLPs) may provide the vestigial signatures through which beyond the Standard Model physics may be first discovered, in particular through exotic decays of the Higgs boson. 
Examples include theories of naturalness, extended Higgs sectors, dark matter, baryogenesis or flavor (see e.g.~\cite{Curtin:2018mvb} and references therein.).
Despite LLPs not being a major design driver for the ATLAS and CMS detectors, they have nevertheless achieved remarkable sensitivity (see e.g.~\cite{CMS-PAS-EXO-16-036,Aaboud:2016dgf,CMS-PAS-EXO-16-003,ATLAS-CONF-2016-103}). On the other hand, there are still important blind spots, some of which can be addressed by LHCb (see e.g.~\cite{Aaij:2016qsm,Aaij:2015tna,Pierce:2017taw,Antusch:2017hhu}) or by beam dump experiments such as NA62~\cite{NA62:2017rwk}. A comprehensive LLP program must however have good sensitivity to  LLPs produced in Higgs decays, something which is notoriously challenging for all of the above experiments. 

The lack of robust coverage for high lifetime LLPs with masses below the weak scale has inspired a number of proposals for dedicated experiments at CERN. The most ambitious along these lines are SHiP~\cite{Alekhin:2015byh} and MATHUSLA~\cite{Chou:2016lxi}. SHiP would be a dedicated beam dump experiment at the Super Proton Synchrotron (SPS), roughly $\sim 150$ m in length, while MATHUSLA would be a detector on the surface above ATLAS or CMS (geometries of $200\,\text{m}\times 200\,\text{m}$, $ 100\,\text{m}\times 100\,\text{m}$ and $ 50\,\text{m}\times 50\,\text{m}$ are being considered \cite{Alpigiani:2631491}). Other proposals take a more opportunistic approach by trading sensitivity for a smaller size and the advantage of being embedded into existing infrastructure. Following this philosophy, MilliQan~\cite{Haas:2014dda,Ball:2016zrp} aims to search for milli-charged particles in a drainage tunnel above CMS; CODEX-b~\cite{Gligorov:2017nwh} proposes to make use of the soon-to-be-vacated data acquisition space next to LHCb; and FASER~\cite{Feng:2017uoz,Feng:2017vli,Kling:2018wct,Feng:2018noy} would consist of a small detector volume in a service tunnel in the far forward regime of the ATLAS interaction point. 

\begin{figure*}[t!]
\includegraphics[width=0.65\textwidth]{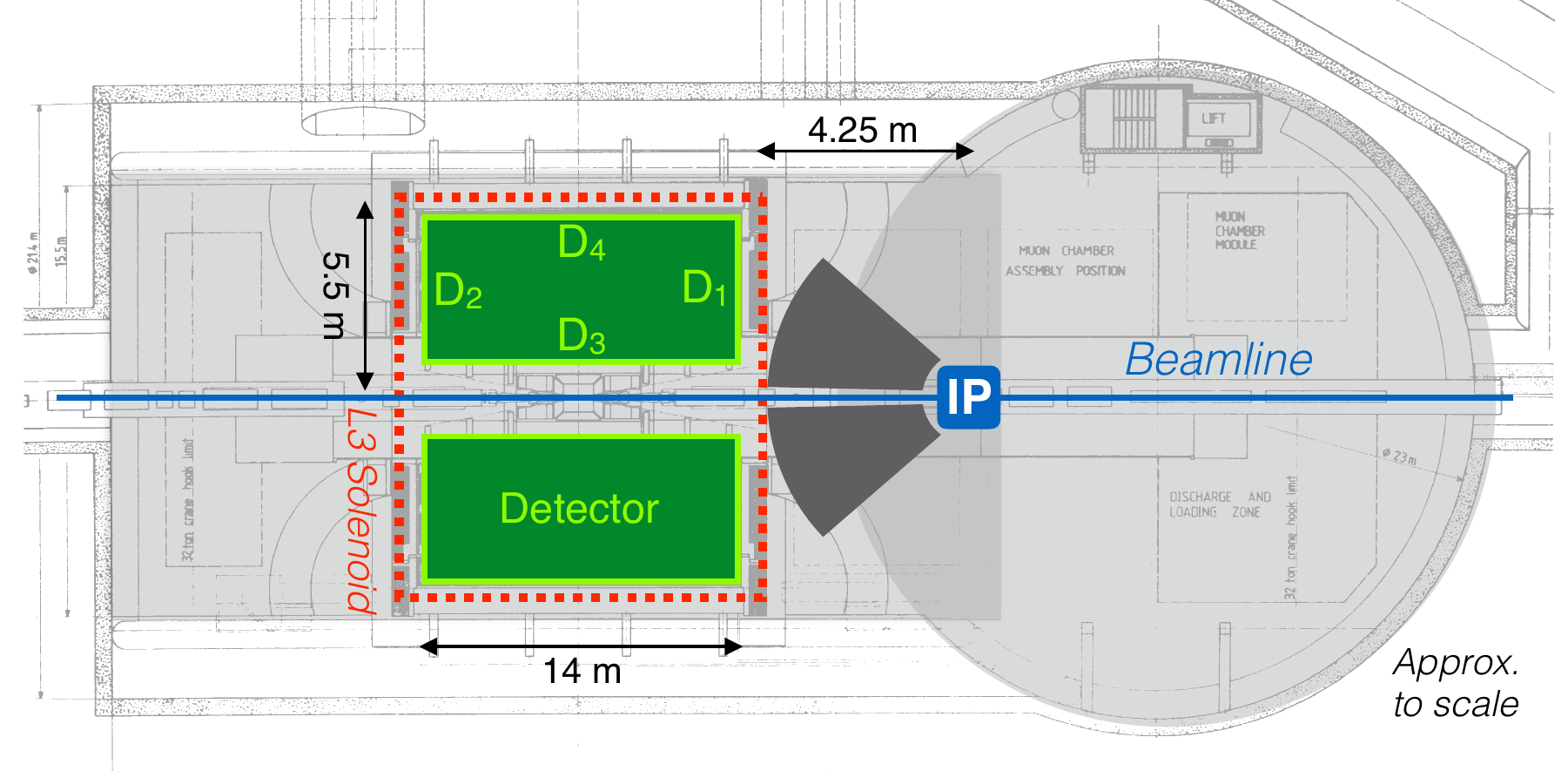}
\caption{Schematic representation of the proposed detector layout.  Cavern layout information is from Ref.~\cite{l3,Aamodt:2008zz}; cavern diagram is reproduced from Ref.~\cite{l3}.  The current L3 magnet is shown in dashed red for reference. The four surfaces bounding the detector volume are labelled D$_{1\ldots 4}$ (see Sec.~\ref{sec:concept} for details).\label{drawing}} 
\end{figure*}

In this paper, we consider an LLP detector constructed inside the L3 magnet that is screened from SM backgrounds by heavily shielding the interaction point, located outside the magnet, as shown in Fig.~\ref{drawing}. 
The proximity of the proposed detector to an LHC interaction point, with a considerable geometric acceptance, permits sensitivity not only to LLPs generated by high center of mass energy portals such as the 
Higgs invisible width, but also from low scale vector, scalar or fermion mixing portals, thereby covering all possible renormalizable couplings of the SM to exotic sectors in one detector concept. 
In this proof-of-concept study, we examine the AL3X reach for an LLP produced in an exotic Higgs or $B$ decay as well as for the production of a kinetically mixed dark photon. 
For an integrated luminosity of order $100~\text{fb}^{-1}$, we find that the AL3X reach meets, exceeds or complements the combined reach of other LLP proposals. 
Much of our discussion will be informed by those applicable to the MATHUSLA~\cite{Chou:2016lxi} and CODEX-b~\cite{Gligorov:2017nwh} proposals, 
though the challenges from backgrounds will be significantly different from the former, and somewhat different from the latter.  

\section{Upgrading IP2\label{sec:challenges}}

Before further motivating and elaborating on the detector concept, we discuss up front some of the potential challenges as they relate to delivering  $\mathcal{O}(100)\text{ fb}^{-1}$ luminosity to IP2 in the AL3X configuration. There are at least four main concerns: (i) moving the IP, (ii) beam quality, (iii) luminosity sharing, and (iv) cost.  

For LHC collisions at 40MHz, an IP can only be moved by multiples of $12.5 \text{ ns} \times c \simeq 3.75$ m.  For this reason we envision moving the IP by $11.25$\,m from its current location at the center of the magnet, 
which should provide sufficient room for shielding the detector from the IP.
However, moving this distance with a reasonably low $\beta^*$ would require changing the layout of the quadrupole magnets in addition to general modifications to the optics.  
The fact that the injection of one of the proton beams is located near IP2 is a possible additional complication.  The current luminosity delivered to IP2 is also so low that it has little impact on the beam quality and lifetime.  
Increasing the instantaneous luminosity to be a non-negligible fraction of the ATLAS and CMS collision rate would make beam preservation more challenging.  
Another consequence of the higher luminosity is that the magnets (triplets and probably also the beam separator magnets) would need additional shielding from forward going radiation.  
In addition to adding absorbers in front of the magnets, one may need to cool the absorbers.   The final concern is the cost.  It is too early to give a reliable price tag of configuring IP2 for AL3X, but given the known feasibility for a similar upgrade at IP8 for LHCb, the cost may not be prohibitive at this time, especially in the context of other ambitious proposals for LLP detectors at CERN. 

At this stage, none of these issues appear to preclude an efficient, robust and cost effective implementation of AL3X, but further engineering studies are required to establish a realistic configuration of the ALICE cavern and the surrounding LHC tunnel and beamline. We further emphasize that the $100\text{ fb}^{-1}$ target is somewhat arbitrary, and is chosen to roughly balance the physics reach against the challenges mentioned above and anticipated limitations from backgrounds. To give the reader a sense of how the various projections scale with the luminosity, we will therefore also show $250\,\text{fb}^{-1}$ projections.
With the above caveats in mind, we now proceed to present the nominal detector concept.

\section{Detector concept\label{sec:concept}}

For LLPs with relatively long lifetimes, the reach of any particular detector scales with the angular coverage and the size of the detector. This is the main reason for the rather large size of the two proposed experiments with the highest sensitivity: SHiP and MATHUSLA.
Since SHiP would operate in beam dump mode off the Super Proton Synchrotron (SPS), the LLPs are necessarily fairly boosted no matter the portal, requiring a long fiducial decay volume.
MATHUSLA would be located $\sim 150$ m from the IP due to space restrictions, and thus
requires a very large detector volume to ensure good geometric coverage. 
Consequently, it is not feasible to instrument MATHUSLA with precision tracking or calorimetry, nor is a magnetic field possible. A similarly ambitious detector, like AL3X, installed near an LHC interaction point, could have the best of both worlds: 
That is, moderately boosted LLPs and access to high center of mass energy  -- e.g.~for Higgs portal production -- but with good geometric coverage in a relatively small fiducial volume. 
Such a more modest volume might be instrumented with a dedicated TPC and potentially a calorimeter, and in the case of IP2, make use of an already existing magnet.\footnote{If IP2 continues to be used for heavy ion physics after Run 4, it may be that the old ALEPH (IP4) or OPAL (IP6) caverns could be used for a proposal similar to what is described in this paper.}

The configuration in Fig.~\ref{drawing} is informed by considerations of both signal acceptance as well as the need to control backgrounds, that is, to look for LLP decays-in-flight in a heavily shielded environment.
The proposed $11.25$\,m shift of IP2 from the center of the L3 magnet provides $4.25$\,m of remaining space that can be used for shielding the IP: 
the L3 magnet half length comprises $6$\,m plus a set of $1$\,m thick solid iron doors, that serve as a return yoke for the magnet. 
As we show in the next section, approximately $40\lambda$ of shielding suffices to suppress the primary hadron and lepton backgrounds to acceptable levels, where $\lambda$ is a nuclear interaction length. 
Although the iron doors provide some shielding already, for the sake of simplicity we will model the shield by $40\lambda$ of tungsten, corresponding to $4$\,m of material. We leave a further optimization of the shield configuration for future work.\footnote{A more realistic and affordable configuration would make use of a tungsten and steel or lead hybrid shield:
In addition to the shielding already provided by the $1$\,m ($6\lambda$) thick iron doors, one could consider  $2.5$\,m ($25\lambda$) of tungsten next to the IP, followed by $1.5$\,m ($9\lambda$) of additional steel or lead. One could also move the IP a further $3.75$\,m away at a mild cost in geometric acceptance, providing enough space for a solely lead or steel shield.}
To veto backgrounds from secondaries produced in the shield itself, an active shield veto is included, embedded inside the shield volume, discussed below. 

The nominal detector geometry is a 12\,m long cylinder, with inner radius $0.85$\,m and outer radius $5$\,m centered on the beamline, 
leaving $\sim1$\,m between the inner (outer) cylindrical detector surface and the beamline (L3 magnet). (This extra space is included 
to allow for support structures and trigger layers, as well as to mitigate some of the forward backgrounds.)
The detector geometry corresponds to a pseudorapidity acceptance $0.9 \le \eta \le 3.7$.

With appropriate shielding of the IP, the occupancy of the detector is expected to be relatively low, even with 40 MHz collisions. A gas TPC could  therefore be a plausible choice for the detector technology because of its excellent tracking resolution, and the possibility of reusing the existing ALICE TPC.  
In our NP sensitivity estimates below, we will consider the reach for the ALICE TPC as well as for a larger TPC filling the entire detector volume. In a realistic design, the size and shape of the volume needed to be instrumented can likely be optimized to an interpolation between these two configurations; we leave this for future studies.

A `time stamp' to enable calibration of the TPC drift time can be achieved by including a trigger layer on the outer surface (D$_4$) and back face (D$_2$) of the cylindrical detector volume, as shown by the light green strips in Fig.~\ref{drawing}. This trigger layer could, e.g., be composed of a scintillator.
The flux of charged tracks, mostly muons, originating from the beamline and the shield is expected to be large. In order to suppress the triggering rate to manageable levels for a TPC readout (1-10 kHz), veto layers on the front (D$_1$) and inner surface (D$_3$) of the detector complement the outer trigger layers, as discussed below in Sec.~\ref{sec:backgrounds}.

Before elaborating on the background estimates of this hypothetical detector, it is worth briefly estimating its fiducial efficiency, as it compares to other proposals. Concretely, in the limit where $\beta\gamma c\tau$ is much larger than the distance of the detector from the IP, the probability for a particle to decay in the detector volume is approximately
\begin{equation}
\epsilon_{\text{fid}} \simeq \frac{\Delta \phi}{2\pi}\int_{\eta_0}^{\eta_1}\!\!\!\! d\eta\, d\gamma\, f(\eta,\gamma)  \frac{\ell}{\beta \gamma c\tau }
\end{equation}
with $\Delta\phi$ the azimuthal angular coverage of the detector, $\eta_0$ ($\eta_1$) the lower (upper) end of the pseudorapidity coverage of the detector, 
$f(\eta,\gamma)$ the distribution of the signal as a function of boost, $\gamma$, and pseudorapidity, $\eta$, and $\ell$ the typical path-length of the LLP in the detector. 
As an example, we compute $f(\eta,\gamma)$ with \mbox{\texttt{Pythia~8.2}~\cite{Sjostrand:2006za,Sjostrand:2014zea}} for an LLP with mass $1$\,GeV produced in an exotic Higgs decay (see Sec.~\ref{sec:reach}),
such that $\gamma \sim \mathcal{O}(100)$. This results in the following comparison
\begin{equation*}
\label{eq:scaling}
\scalebox{0.8}{
\parbox{1.3\textwidth}{
\newcolumntype{C}{ >{\raggedright\arraybackslash $} l <{$}}
\begin{tabular}{lCCC}
	\text{AL3X}: 		&	0.9<\eta<3.7;	&\dfrac{\Delta\phi}{2\pi}=	1;	& 	\epsilon_{\text{fid}}  =  \dfrac{3.2\times 10^{-2}}{c\tau/\text{m}}\,,\\
	\text{MATHUSLA}: 	&	0.9<\eta<1.4;	&\dfrac{\Delta\phi}{2\pi}=	0.15;	&	 \epsilon_{\text{fid}} = \dfrac{6.9\times 10^{-3}}{c\tau/\text{m}}\,,\\
	\text{CODEX-b}: 	&	0.2<\eta<0.6;	&\dfrac{\Delta\phi}{2\pi}= 0.06;	& 	\epsilon_{\text{fid}} =  \dfrac{1.1\times 10^{-3}}{c\tau/\text{m}}.\\
\end{tabular}
}}
\end{equation*}
Being the closest the IP, AL3X has the largest angular coverage off all proposals, though the typical path-length of an LLP is a factor of $\sim 2$ less than in MATHUSLA ($\sim12$~m vs $\sim 25$~m). 
Since AL3X is more forward than MATHUSLA and CODEX-b, the LLPs tend to be a bit more boosted as well. 
In the long-lifetime limit, AL3X ends up having somewhat larger fiducial efficiency than MATHUSLA, for which we  have assumed the 100~m $\times$ 100~m configuration \cite{Alpigiani:2631491}. 
The efficiency for MATHUSLA's 200~m $\times$ 200~m configuration is roughly a factor of 3 larger, and similar to that of AL3X. The relative sensitivity between both detector concepts will therefore largely be driven by the luminosity that could be delivered to IP2 in Run 5.  

In the short life-time regime, this scaling does not apply and instead the main driver of the sensitivity is the distance of the IP to the detector, in the frame of the LLP. 
As we will see in Sec.~\ref{sec:darkphoton}, for a kinetically mixed dark photon, an experiment like AL3X could be competitive with SeaQuest, FASER and SHiP in this regime, despite the lower number of collisions. 
The reason is the relatively short baseline of AL3X, as compared to FASER, and its access to very highly boosted LLPs, as compared to SeaQuest and SHiP.

\section{Backgrounds and Shielding}\label{sec:backgrounds}
There are two classes of backgrounds for AL3X: Those which are attenuated by the shield and those that are not. We discuss both in turn, as well as the necessary specifications for the shield to achieve the desired low background regime. The shielding analysis is similar in spirit to the CODEX-b proposal~\cite{Gligorov:2017nwh}, though the background analysis and shield design for AL3X is complicated by the detector surrounding the beamline itself.

\subsection{Shield configuration}
\begin{figure}[t]
\includegraphics[width=0.5\textwidth]{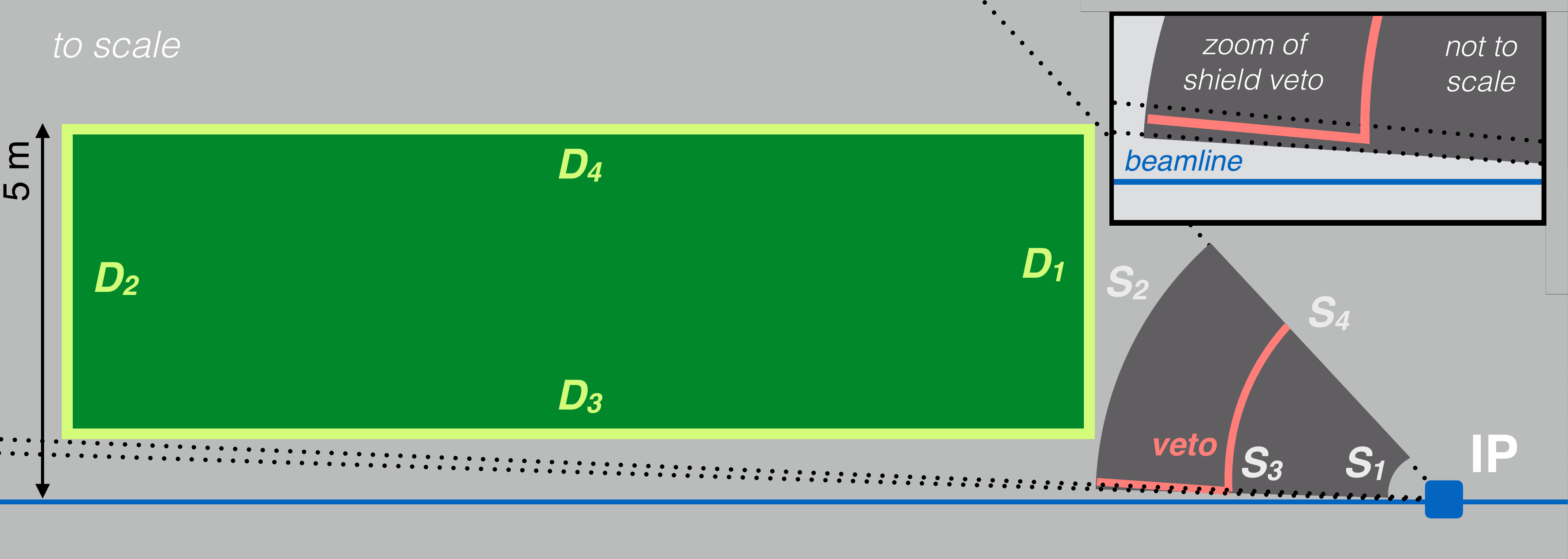}
\caption{Configuration of the shield, to scale. }
\label{fig:shield}
\end{figure}

The proof-of-concept background shield configuration is taken to be a $40\lambda$ ($4$\,m) spherical shell segment of tungsten, 
centered on the IP with an inner and outer radius of $0.25$\,m and $4.25$\,m respectively, covering an pseudorapidity range $0.8 < \eta < 4.3$, as shown in Fig.~\ref{fig:shield}.
As for the CODEX-b shield, the prompt primary background fluxes pass through the S$_1$ surface and are then attenuated by the shield. The primary fluxes, in particular muons, may produce secondary backgrounds via scattering inside the shield. These secondary backgrounds can be reduced by a judiciously located active veto inside the shield itself. 
However, the extremely large forward production of pions near to the beamline means that the cavity around the beamline itself can also be a source of a large flux of daughter muons, 
that instead transect the S$_3$ inner surface  -- `shield-clipping' muons -- producing copious secondary neutrons and kaons, or miss the shield entirely -- `shield-evading' muons.
To control these backgrounds, the shield coverage is extended beyond the angular acceptance of the detector, and an additional, radially oriented active veto is included, 
as shown in Fig.~\ref{fig:shield}. Except when explicitly stated below, effects of shield-clipping muons are found to always be highly subleading compared to the background fluxes from muons traversing the full shield, and are therefore hereafter neglected.

Control of the background processes in AL3X are determined by detector and signal-specific interplay between three different rates: 
\begin{enumerate}
	\item[(i)] the detector trigger rate;\\[-15pt]
	\item[(ii)] the shield veto rate;\\[-15pt]
	\item[(iii)] and the `potentially irreducible' background rate.
\end{enumerate}
The first of these is limited by the capabilities of the TPC readout, while being driven by the total charged flux through the detector, that can be large. 
An appropriate triggering strategy will reduce this rate to acceptable levels. 

The shield veto rate is driven by the requirement that neutral secondary backgrounds 
produced downstream in the shield -- mainly muoproduction of $K_L$'s and neutrons -- can be vetoed down to acceptable levels by tagging the primary muons. 
This veto rate must not be so large that a significant fraction of all events are vetoed. 
The trigger rate sets an upper bound for the rate at which the shield veto needs to be read out, so that a fast shield veto readout need not be required.

Finally, the irreducible background rate sets the sensitivity to low rate signals. These backgrounds may arise mainly from the abovementioned secondary $K_L$'s and neutrons or primary neutrinos.
The extent of the detector's ability to reduce these backgrounds is both detector technology and LLP signal dependent: signals with no missing energy will be easier to distinguish 
from these backgrounds than missing energy LLP decay signatures, since one can require the vertex to point back to the IP. (This will be the case for all signal benchmarks we consider in Sec.~\ref{sec:reach}.)
It is likely that signals and backgrounds can be somewhat well-characterized and separated by using the TPC and $B$-field of the L3 magnet, 
though a full examination of these capabilities is beyond the scope of this proof-of-concept study. We therefore refer to the remaining backgrounds as `potentially irreducible' rather than irreducible. 
To eliminate some of the backgrounds which are soft and more difficult to estimate, we will hereafter impose a $3$\,GeV cut on the scalar sum of the momenta of the tracks of a candidate vertex. The effect of this cut on the signal efficiency for the benchmark models considered in Sec.~\ref{sec:reach} is 10\% or less.

\begin{table*}[t]
\renewcommand*{\arraystretch}{1.3}
\newcolumntype{C}{ >{\centering\arraybackslash $} m{2.7cm} <{$}}
\newcolumntype{D}{ >{\centering\arraybackslash $} l <{$}}
\newcolumntype{E}{ >{\centering\arraybackslash $} c <{$}}
\begin{tabular*}{0.95\linewidth}{@{\extracolsep{\fill}}D|CC|C|C|E}
\hline
\multirow{3}{*}{BG species} & \multicolumn{2}{c|}{Full shield (S$_{1}$--S$_{2}$)} & \text{Evade shield} & \multirow{3}{*}{\shortstack{Net BG flux/pp \\ into detector \\ (no cuts)}} & \multirow{3}{*}{\shortstack{BG rate \\ per $100$\,fb$^{-1}$}} \\ 
 & & & & & \\
  & \text{shield veto rate}  & \text{BG flux/pp }  &\text{BG flux/pp} & & \\
 \hline \hline
 	 n+\bar n~(>3\,\text{GeV})      		& \text{---}   		& 4. \times 10^{-16}  		 		&   \text{---}      	& 3. \times 10^{-6}   &  \lesssim 0.2 \\
 	 p+\bar p        					&  2. \times 10^{-6}	& 1. \times 10^{-14}  				&  \text{---}      	& 5. \times 10^{-7} &  \text{---}  \\
	 \mu            					&  0.006			& 3.\times 10^{-11}  		 		&   0.007  		& 0.01 &  \text{---}  \\
	 e              					&  5. \times 10^{-7}	& 3. \times 10^{-15}  		  		&   \text{---}     	& 3. \times 10^{-7} &  \text{---}  \\
	 K^0_L         					& \text{---}  		& 1. \times 10^{-15}  		   		&   \text{---}     	& 6. \times 10^{-8} &  \lesssim 1 \\
	 K^0_S  						& \text{---}  		& 4. \times 10^{-16}  		  		&   \text{---}      	& 3. \times 10^{-8} &  \ll 1 \\
	 \gamma  	    					& \text{---}   		& 1. \times 10^{-15}  		  		&   \text{---}      	& 1. \times 10^{-7} &  \text{---}  \\
 	 \pi^\pm        					& 2. \times 10^{-6}	& 5. \times 10^{-15}  		  		&   \text{---}      	& 4. \times 10^{-7} &  \text{---}  \\
	 K^\pm        					& 2. \times 10^{-7}	& 9. \times 10^{-16}  		  		&   \text{---}      	& 8. \times 10^{-8} &  \text{---}  \\
 	 \nu+\bar\nu~(>3\,\text{GeV})  	& \text{---}  		& 0.01 	            		    		&   3. \times 10^{-4}   & 0.2  &  \lesssim 10  \\
 \hline
\end{tabular*}
\caption{Results from the preliminary \texttt{Geant4} background simulation for $(20 + 20)\lambda$\,W shield, i.e. with an active shield veto at $20\lambda$, applying a veto efficiency of $\epsilon = 10^{-8}$. For outgoing neutrons and neutrinos a cut on their kinetic energy was applied, as indicated in the first column. Background (BG) fluxes per pp collision (pp) are shown for fluxes entering the detector by traversing the full shield (S$_{1}$--S$_{2}$) or by missing the shield entirely (Evade), together with veto rate for charged BG fluxes passing through the veto itself. Also shown are (upper bounds for) the net background fluxes that enter the detector, i.e. without the application of the veto rejection factor, relevant for the trigger veto rate.  Actual potential background rates for $100$/fb, shown in the final column, are obtained from the BG fluxes/pp, by folding in the decay or scattering probabilities, which are detector dependent, and assuming a minimum bias cross-section of $100$\,mb (see text for details).}
\label{tab:bkg-fluxes}
\end{table*}

\subsection{Shield-attenuated backgrounds}

To estimate the backgrounds, we simulate minimum bias production of pion, kaon, muon, neutron, proton and neutrino fluxes with \texttt{Pythia~8}~\cite{Sjostrand:2006za,Sjostrand:2014zea}. 
Leptons produced from pion decay vertices at $\eta < 0.8$ and $r > 0.75$\,m are neglected, under the assumption that their parent pions can be suppressed with a moderate amount additional shielding close to the IP, external to the geometric acceptance of the primary shield.

The propagation of primary backgrounds and production of secondary backgrounds inside tungsten is simulated with \texttt{Geant4} 10.3 with the \texttt{Shielding}~2.1 physics list for which high energy interactions are modeled with the \texttt{FTFP\_BERT} physics list 
based on the Fritiof~\cite{ANDERSSON1987289,Andersson1996,NilssonAlmqvist:1986rx,Ganhuyag:1997gz} and Bertini intra-nuclear cascade~\cite{Guthrie:1968ue,Bertini:1971xb,Karmanov:1979if} models 
and the standard electromagnetic physics package~\cite{1462617}.  
Propagating a large amount of events though the full shield is computationally prohibitive, so we instead use a ``particle gun'', binned in energy and particle species, applied to a 5$\lambda$ shield subelement (see App.~\ref{sec:grids}). 
The resulting map between the incoming and outgoing fluxes is then applied recursively to obtain the attenuation and response of the full $40\lambda$ shield.  
Neutrino production of neutral hadrons occurs at a prohibitively small rate and is not included in this analysis; these backgrounds are discussed in Sec.~\ref{sec:otherbg} below.

An active veto layer is located at a depth of $20\lambda$ inside the shield, with a rejection factor $\epsilon = 10^{-8}$, achievable e.g.~with several redundant layers of scintillator. 
The purpose of this `shield veto' is to detect charged tracks that may produce neutral secondary fluxes 
-- $K_L$'s or neutrons -- downstream in the shield itself, that may then enter the detector and produce an LLP-like event by decay or scattering. 
The location of the veto is determined by a balance between detecting charged particles before they create secondaries, not having too large a shield veto rate, and having sufficient material downstream of the veto to suppress neutral primary or secondary fluxes through the veto.
The expected correlation between primary charged fluxes and neutral secondary fluxes within the shield -- when a charged particle produces a secondary, it is typically not fully stopped -- in principle permits vetoing some of these neutral secondaries produced upstream of the veto layer, so that the veto might be located deeper in the shield with a correspondingly lower shield veto rate. To be conservative, we have assumed the charged and neutral fluxes are instead fully decorrelated. 
The corresponding shield veto rate derived from this analysis, as well as the amount of shielding material required, is therefore expected to be an overestimate.

\begin{figure*}[t]
	\includegraphics[width=0.32\textwidth]{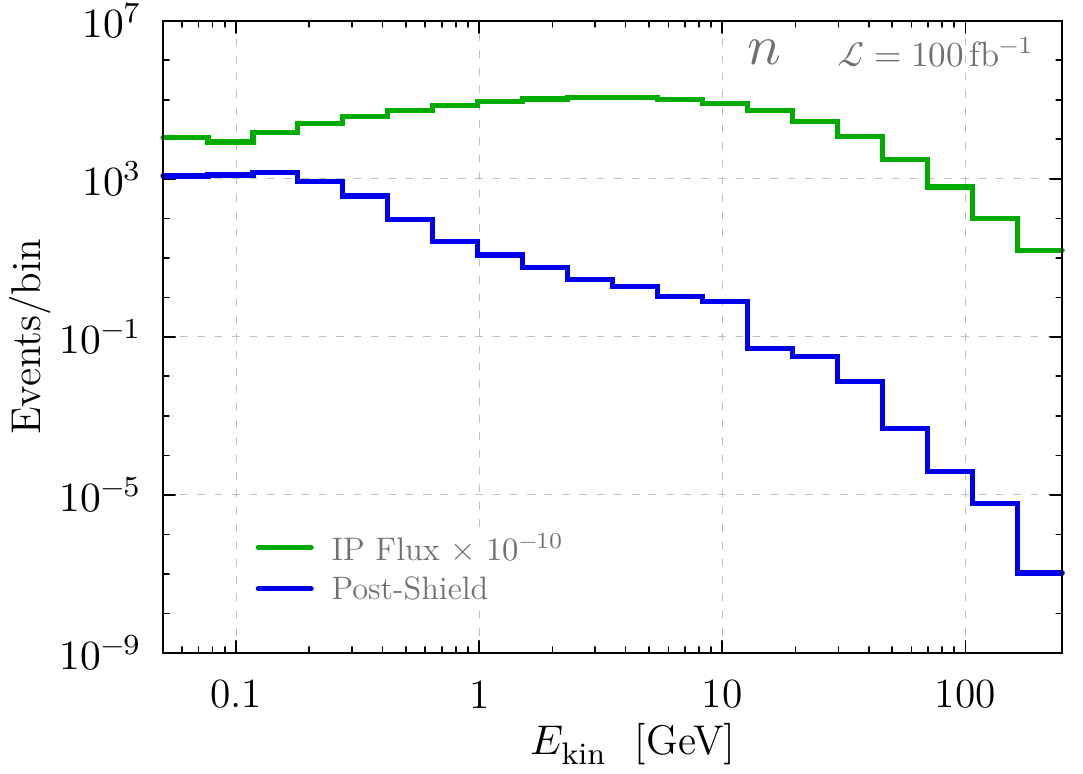}\hfill
	\includegraphics[width=0.32\textwidth]{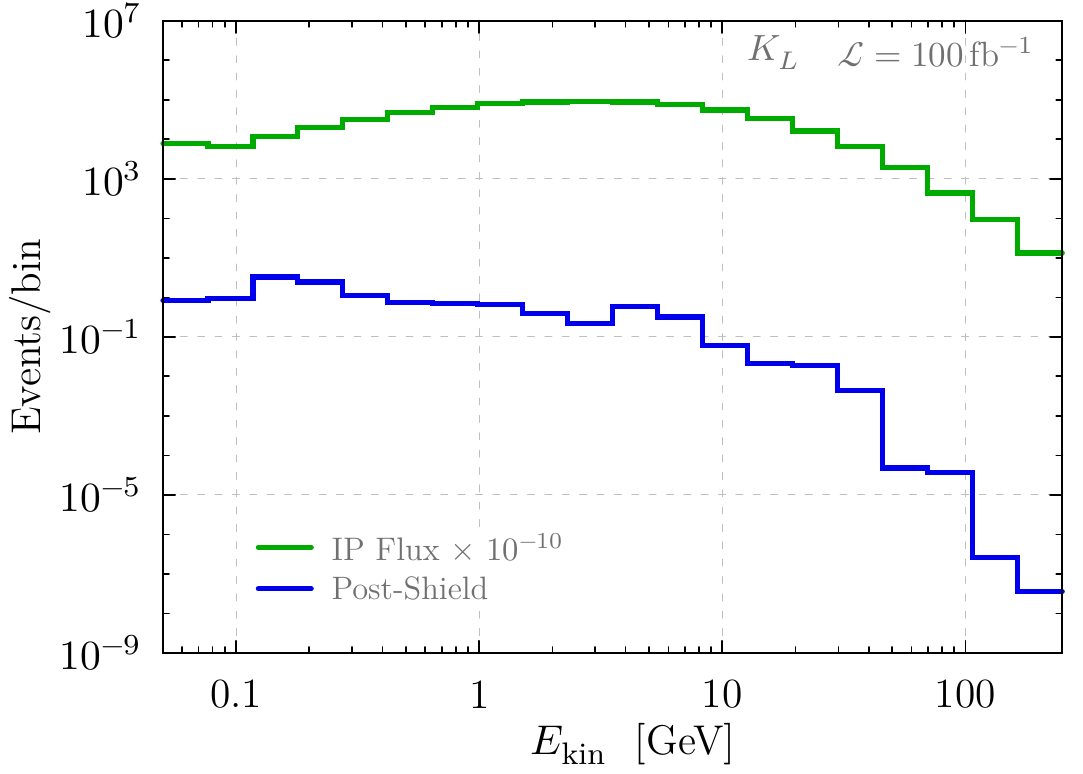}\hfill
	\includegraphics[width=0.32\textwidth]{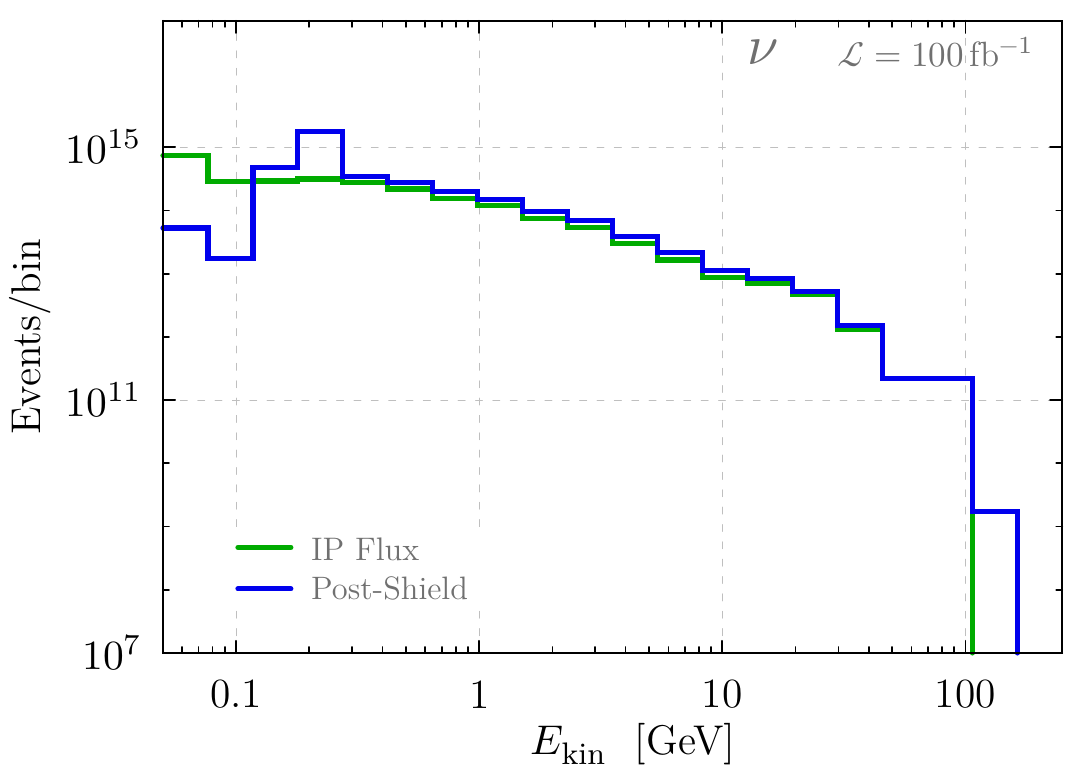}
\caption{Kinetic energy spectra of outgoing neutrons, $K_L$'s and neutrinos (blue), compared to their fluxes from the IP (green). For the sake of visual clarity, both the neutron and $K_L$ fluxes from the IP are scaled by a factor of $10^{-10}$. }
\label{fig:BGflx}
\end{figure*}

In Table~\ref{tab:bkg-fluxes} we show the efficiencies (background flux/pp collision) for each relevant primary and secondary background 
entering the detector volume after propagation through the shield and application of the shield veto, integrated over energy above a minimal threshold. 
Also shown are the `shield veto rates', corresponding to the flux of charged particles through the veto itself, relevant for an estimate of the event rejection rate by the shield veto. 
We divide the background fluxes, as appropriate, into those that transit the full shield, i.e. S$_{1}$--S$_{2}$, and those that are produced by shield-evading muons. 
Effects of shield clipping muons are negligible for all backgrounds, with the exception of the muon rate itself, for which they comprise approximately $50\%$ of the S$_{1}$--S$_{2}$ rate, i.e. a flux/pp collision of $0.3\%$.
We also provide, for the purpose of estimating the maximum required detector-trigger rate, the net background flux into the detector volume after propagation through the shield, 
but without application of the shield veto, in the second-to-last column. In order to characterize the sensitivity of the background rates to the $3$\,GeV cut, in Fig.~\ref{fig:BGflx} we show the neutron, $K_L$ and neutrino kinetic energy spectra.

The shield veto rate is driven mainly by the muon flux, and is approximately $\sim (0.6 + 0.3)$\%. Including a pile-up factor of $\sim 10$, this implies a $10\%$ event rejection 
rate by the shield veto.  
As discussed above, this event veto rate is an overestimate, expected to be reduced once correlations between charged and neutral fluxes in the shield are included. 
There is also likely a spatial correlation between the charged primary and neutral secondary velocities, such that detector and veto might be segmented. 
This would permit vetoing only part of the detector, with the remainder open to detect a signal.

The net background flux into the detector is overwhelmingly dominated by muons, which have a flux/event of $\sim (0.6 + 0.3 +0.7)\% \times$pile-up. While this background can easily be eliminated by detecting a track in the front of the detector, it can contribute to the trigger rate.
In particular, muons would induce a trigger rate on D$_2$ and D$_4$ of $\mathcal{O}(\text{MHz})$, which is far too high for the TPC readout. 
However, one can use all four trigger/veto layers on the detector surfaces to significantly reduce the trigger rate from muons, by computing the number of hits in D$_2$ and D$_4$ less the number of hits in D$_1$ and D$_3$:
\begin{equation}
	T  = \#\,\text{hits}(\text{D}_2 + \text{D}_4) - \#\,\text{hits}(\text{D}_1 + \text{D}_3).
\end{equation}
Nominally $T=0$ for background muons and $T>0$ for an LLP decay, up to small instrumentation inefficiencies:  
The probability of tracks from an LLP decay hitting D$_1$ or D$_3$ is low, since they are expected to be mostly forward moving, and the average projective size of D$_3$ is only $\sim 10\%$ of D$_4$.
A triggering strategy that requires $T>0$ may then reduce the trigger rate to acceptable levels. If needed, one may further segment the $T$ variable azimuthally. 
The flux of other backgrounds will also contribute to the trigger rate, either as charged particles directly transiting the trigger layers -- i.e.~the protons, electrons and pions -- 
or by decaying or scattering to charged tracks in the detector volume -- i.e.~from the kaons and neutrons. However, these induce trigger rates of $\mathcal{O}(\text{Hz})$, well within TPC readout capabilities.  Finally, cosmogenic muons would also trigger the detector with a rate of $\mathcal{O}(\text{kHz})$ \cite{Drollinger:883295}, though if needed this rate can be reduced substantially with a timing cut.

Folding the post-shield background fluxes against their relevant decay or scattering probabilities allows estimation of the total potential neutral background rates into the detector volume, shown in the final column of Table~\ref{tab:bkg-fluxes}.
These are estimated assuming a minimum bias cross-section of $100$\,mb, and for the estimation of scattering probabilities, we assume most of the TPC gas is Neon at standard temperature and pressure. For the neutron background, only events which produce at least two tracks are relevant, notably reactions of the form $ n n \to n n \pi^+\pi^-$ and $n n \to n p^+ \pi^-$.  Using isospin symmetry, we estimate both processes from the analogous $pp$ reactions \cite{smellybook}, which results in a combined rate of $\sim 5$ mb per nucleon for $E_{\text{neutron}}\gtrsim 3$ GeV. Accounting for the TPC's size and gas pressure, this implies that roughly 5\% of the neutrons entering the detector will create two or more tracks. This leads to an upper bound of $\sim0.2$ neutron-induced background events per $100$\,fb$^{-1}$. It should be emphasized that this is still a substantial overestimate of the actual background rate, since these reactions predominantly occur at very low momentum transfer (see e.g.~\cite{PhysRev.161.1387}), resulting in mostly soft tracks. In reality the neutron must thus be substantially harder than 3 GeV for the tracks to satisfy our 3 GeV cut on the scalar sum of their momenta, further reducing the background rate.  
We also note that a large control sample for this background can be acquired by simply inverting the shield veto, and it should be possible to characterize it carefully with data-driven methods. Moreover, the ongoing data-driven calibration for the CODEX-b shield will likely assist in calibration of expected background rates at AL3X.

\subsection{Other backgrounds}
\label{sec:otherbg}

Backgrounds induced by atmospheric neutrinos have been shown to be negligible for a detector of this size~\cite{Gligorov:2017nwh}, but neutrinos originating from decays of primary pions must be considered: the kinetic energy spectrum of primary and secondary neutrinos entering the detector is shown in Fig.~\ref{fig:BGflx}. 
We have verified that the contribution from pion decays dominates that of prompt neutrino production from $c\bar c$ and $b\bar b$ production by roughly an order of magnitude. The main process of concern is either coherent particle production of the neutrinos on the gas of the TPC, e.g.  $\nu n \to \mu p \pi$, $\nu n \to p \mu$, or deep inelastic scattering. We require the neutrino to have energy $E_\nu \ge 3$\,GeV, and conservatively assume that all the remaining scattering events satisfy our 3 GeV cut on the scalar sum of the track momenta. The  neutrino-nucleon cross section is taken from Ref.~\cite{Formaggio:2013kya}; 
as a rough guide, for neutrinos with $E_\nu \gtrsim 1$\,GeV, the cross-section per nucleon $\sim  0.01 (E_\nu/\text{GeV})$\,pb. 
Integrating the cross-section against the primary neutrino flux implies $\mathcal{O}(1)$ $\nu$-Ne scattering events per $100$\,fb$^{-1}$. This background is threshold 
dominated~\cite{Rodrigues:2015hik}, which means that for a charged (neutral) current interaction, the majority of the energy is carried away by a relatively 
hard muon (neutrino), accompanied with a few soft hadronic tracks, that do not point to the IP.  It should therefore be possible to further reduce this background by
placing additional cuts on the minimal track momentum or by applying an impact parameter cut. 

The neutrino flux may also interact inelastically with the shield and create secondary neutral hadrons, in particular neutrons or $K_L$'s. If this occurs in the last few interaction lengths, these secondary hadrons may reach the detector. The MINERvA collaboration measured the inclusive, neutral-current $K^+$ production cross-section for a neutrino beam centered around $E_\nu\approx 3.5$ GeV to be $\sim 10^{-40}\,\text{cm}^{2}$ \cite{Marshall:2016yho}, which we take as an approximation for the $K_L$ production cross section. Charged current scatters can be vetoed effectively by tagging the associated muon, but we do not exploit this here. Since the neutrino-nucleon cross section is expected to rise linearly with energy, we weight the MINERvA result with $E_\nu/\text{3.5 GeV}$ and fold this against the neutrino spectrum obtained from \texttt{Pythia~8} (described above).  As a result we find $\mathcal{O}(3)$ $K_L$'s which are produced in the last interaction length of the shield. These residual $K_L$ should be further attenuated by an $\mathcal{O}(1)$ number under propagation through the last interaction length. Since they are expected to be soft, they are further likely reducible by a cut on track momenta and/or a requirement that the tracks point to the IP.

Using the total inelastic cross section~\cite{Formaggio:2013kya}, we bound the amount of neutrons produced in the last interaction length of the shield to be less than $\mathcal{O}(300)$. With a $\sim 5\%$ scattering probability in the TPC gas (see previous section), this implies an upper bound of $\mathcal{O}(15)$ events. This number is conservative in several ways: (i) The produced neutrons should substantially lower energy than the incoming neutrino, further softening the spectrum and (ii) in the majority of the events one or more charged states will be created along with the neutron. These charged states are likely to reach the detector as well and can therefore be used to veto the event. 
A full simulation, including a realistic detector response, is beyond the scope of this work, but for the time being it appears plausible that these handful of background  events can be fully reduced by the kinematic cuts described above. 

Finally, there are additional sources of background, such as cavern backgrounds, beam-gas backgrounds, hot-shield-induced backgrounds.  We expect that none of these are a major source of background, but they would need to be carefully considered, possibly with in-situ studies, should the AL3X proposal move forward. The cavern and hot-shield-induced backgrounds are likely eliminated by the cuts on the track momenta, but are expected to contribute to the noise levels in the detector. Beam gas events can produce neutral hadrons which may bypass the shield. These events are very boosted in the lab frame, and for the tracks to hit the TPC, the beam-gas vertex must be either be located far behind the IP, in which case the hadrons would pass through the full shield, or in the far forward region of the beamline well beyond the detector volume. In the latter case, any vertex made in the detector would generate tracks pointing towards, rather than from, the IP, which will not occur for an LLP vertex.  A hadron from a beam gas event could in principle also deflect off the beam pipe or create more secondaries, and for a realistic design it may be therefore be adviseable to clad the beampipe with a layer of tungsten, as is the case for the current forward muon detector of ALICE \cite{Aamodt:2008zz}. The beam-induced backgrounds have been measured in the hotter ATLAS cavern and were found to be small ($\ll 40$ MHz)~\cite{Aad:2016tzx}, and should not affect the trigger strategy laid out in the previous section. 

In summary, our baseline configuration has a trigger rate, veto rate, and irreducible background rate that are compatible with a close to zero background search for $100$\,fb$^{-1}$ of luminosity.  A trigger rate well-below the TPC bandwidth can be achieved with a relatively simple algorithm using the $T$ trigger variable, while the fraction of events vetoed by the shield is $\sim 1\%\times$pile-up.  This veto can be applied offline, as the detector trigger rate is sufficiently slow.  

\section{Reach}
\label{sec:reach}
In this section we present three example benchmark models, representing high, medium and low mass portals. For our reach estimates we require three signal events, which roughly corresponds to a 95\%CL exclusion, assuming zero background.

\subsection{Exotic Higgs decays}
\label{sec:Higgs}

Searching for exotic Higgs decays are a top priority for the HL-LHC program. The small width of the SM Higgs means that relatively small couplings to exotic states, with mass $< m_h/2$, 
can lead to an appreciable Higgs exotic branching ratio. Combined with the large sample of Higgs bosons expected from HL-LHC -- approximately $10^8$ Higgs -- this leads to a powerful portal for probing new physics.
As a benchmark, we consider an exotic decay of $h\to XX$, with $X$ a long lived particle decaying to two or more charged SM particles. $X$ could for instance be a kinetically mixed dark photon 
(e.g.~\cite{Schabinger:2005ei,Gopalakrishna:2008dv,Curtin:2014cca,Strassler:2008bv}) or another (pseudo-)scalar of an extended Higgs sector (e.g.~\cite{Curtin:2013fra}). 

\begin{figure}\centering
\includegraphics[width=0.5\textwidth]{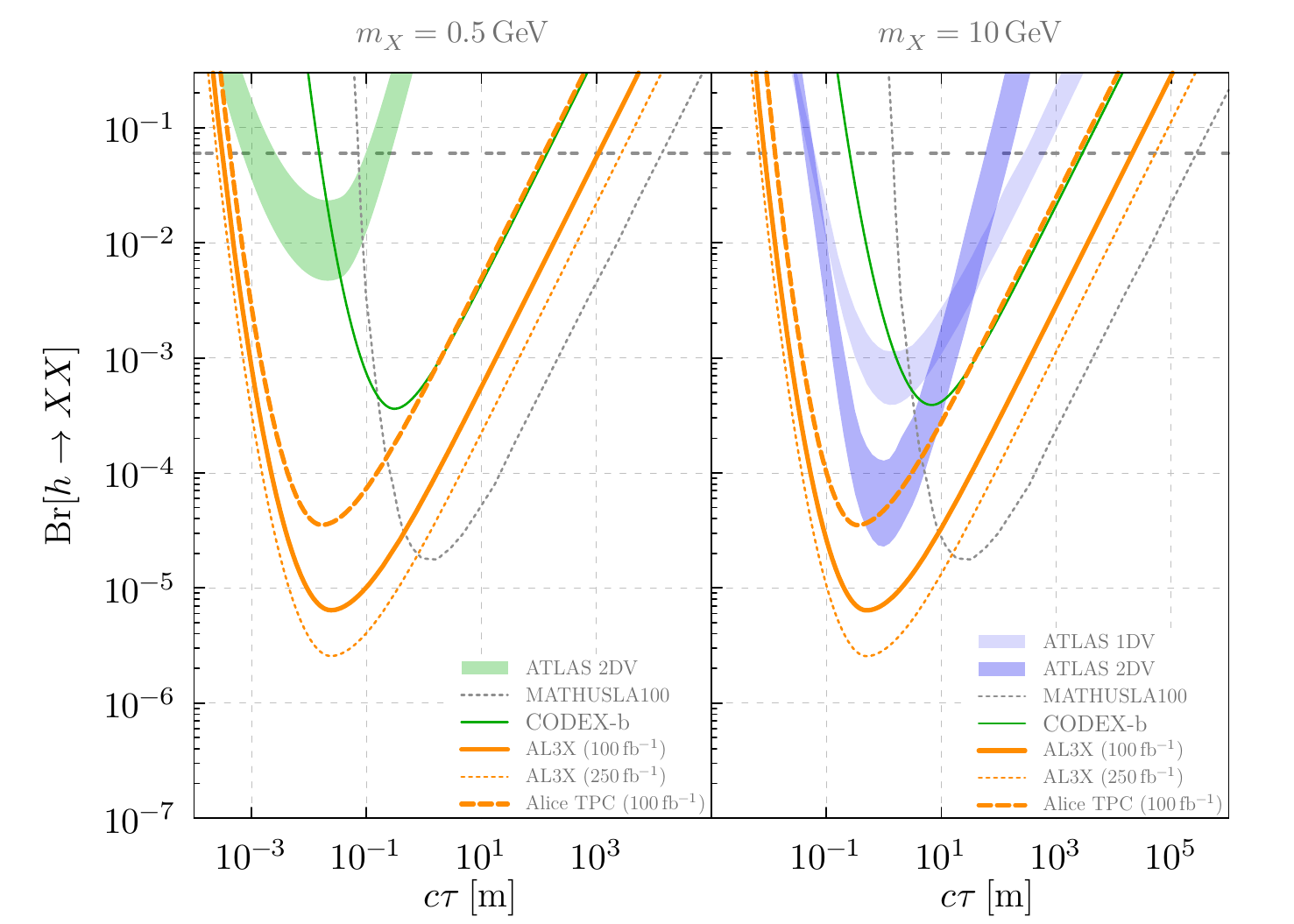}
\caption{Projected reach for AL3X, CODEX-b, MATHUSLA and ATLAS (see text) for $h\to XX$. For MATHUSLA, the $100\,\text{m}\,\times\, 100\,\text{m}$ configuration was assumed~\cite{Alpigiani:2631491}. The reach for $h \to \text{invisibles}$ is also shown (horizontal gray dashed)~\cite{CMS:2013xfa}. \label{fig:Hdecay}}
\end{figure}

We estimate the fiducial efficiency for this benchmark with \texttt{Pythia~8}, and show the resulting reach for 95\% exclusion in Fig.~\ref{fig:Hdecay}, assuming negligible irreducible backgrounds. 
We see that AL3X can reach $h\to XX$ branching ratios as low as $\sim 5\times 10^{-6}$, which is close to the best reach that is achievable with a $100\,\text{fb}^{-1}$ data sample, 
corresponding to $6\times 10^6$ Higgs bosons. 
In the large lifetime limit, the AL3X reach falls in between the reach for \mbox{CODEX-b} and \mbox{MATHUSLA}. 
For comparison, we also show the (optimistic) reach for 250 $\textrm{fb}^{-1}$ and the reach assuming only the existing ALICE TPC for $100\,\text{fb}^{-1}$.  Naturally, AL3X would have much better reach at low c$\tau$ regardless of the delivered luminosity, 
since it is much closer to the IP.

The projected ATLAS reach for a Higgs decay to a pair of displaced dijets is also included in Fig.~\ref{fig:Hdecay}: the shaded bands attempt to indicate the uncertainty on these projections.  For ATLAS and CMS, the mass of the LLP is a crucial parameter, as the number of tracks associated with the vertex is a key discriminant between signal and background. This is why the ATLAS reach for the $0.5$\,GeV benchmark~\cite{ATLAS-CONF-2016-042} in Fig.~\ref{fig:Hdecay} is substantially less than for the 10 GeV mass benchmark ~\cite{Aad:2015uaa,Coccaro:2016lnz}: The lower edge of the green ATLAS band is obtained by rescaling the current expected limit in~\cite{ATLAS-CONF-2016-042}, assuming that the systematic uncertainties could be lowered with a factor of five (For more details, see Sec.~III of Ref.~\cite{Gligorov:2017nwh} and see \cite{Chatrchyan:2012jna,CMS:2016tgd} for the analogous searches by CMS.) To further reduce the backgrounds, ATLAS and CMS often require two displaced vertices, as indicated by the ``2DV'' label in Fig.~\ref{fig:Hdecay}, as compared to higher background searches with a single displaced vertex (``1DV''). In the latter case, both the current and future limits are merely projections~\cite{Coccaro:2016lnz}, and it is conceivable that innovations in future analyses may substantially improve on this. For these reasons, largest increase in sensitivity from proposals such as AL3X, CODEX-b and MATHUSLA over ATLAS or CMS will be for LLP's with mass $\lesssim 10$ GeV, though there is gain for higher mass LLPs as well. Especially for low mass LLP's, the reach on the Higgs branching ratio can be improved by several orders of magnitude.

\subsection{Exotic \texorpdfstring{$B$}{B} decays}
\label{sec:bdecays}

A new scalar state $\X$, lighter than the $B$ meson, can also be produced in a neutral current $B \to X_s \X $  decay, even if the coupling of $\X$ to the SM satisfies the ansatz of minimal flavor violation. A canonical example is the case where $\X$ mixes with the SM Higgs, and thus obtains a coupling to the SM fermions proportional to their masses. The inclusive branching ratio for this process is~\cite{Willey:1982dk,Chivukula:1988lo,Grinstein:1988yu}
\begin{align}
	\label{eqn:BRIC}\nonumber
	\frac{\text{Br}[B \to X_s \X]}{\text{Br}[B \to X_c e \nu]}  &\simeq \frac{27 g^2 s_\theta^2}{256 \pi^2} \frac{m_t^4}{m_b^2 m_W^2} \frac{\lambda^2\big(\frac{m_\varphi}{m_B},\frac{m_{K}}{m_B}\big)}{f(m_c/m_b)} \bigg|\frac{V_{ts} V_{tb}}{V_{cb}}\bigg|^2\,\\
	&\approx 6. \times \lambda^2\big(m_\varphi/m_B,m_{K}/m_B\big)\,,\\
\intertext{where}
\lambda(x,y)&\equiv\sqrt{\big(1-(x-y)^2\big)\big(1-(x+y)^2\big)}\,,
\end{align}
and $s_\theta\equiv\sin\theta$ parametrizes the Higgs-$\X$ mixing.

The $\X$ must decay back to the SM through its induced Yukawa couplings with the SM fermions. Its lifetime is therefore also determined by $s_\theta^2$, but is affected substantially by hadronic resonances for $m_\X \gtrsim 1$\,GeV, as well as threshold effects. The theory uncertainties in this region are rather large, and we make use of the data-driven result from Refs~\cite{Fradette:2017sdd,Bezrukov:2009yw}. (This result is in good agreement with another, more recent calculation of the lifetime \cite{Winkler:2018qyg}.) For our estimates we assume a $b\bar{b}$ production cross-section of $500$\,$\mu$b and compute the boost and pseudo-rapidity distributions with \texttt{Pythia~8}. 

\begin{figure*}
\includegraphics[width=0.45\textwidth]{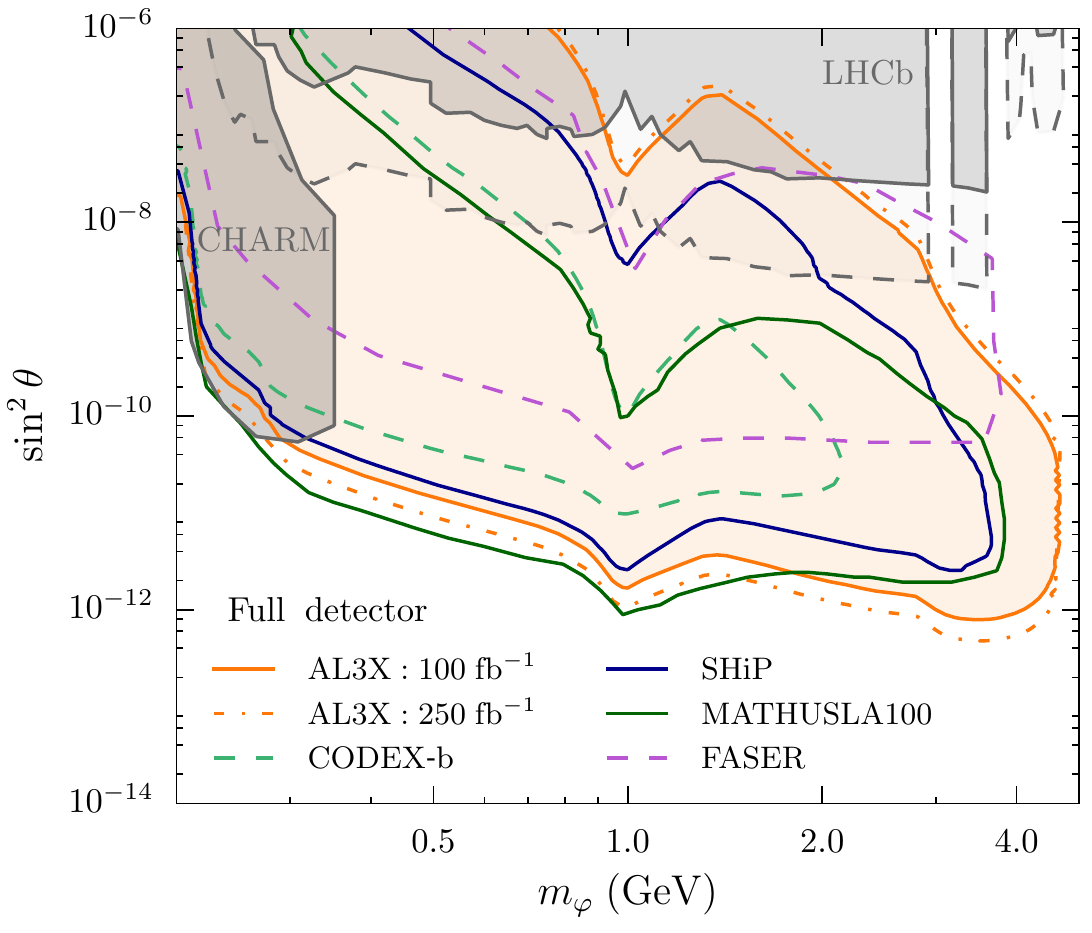}\hfill
\includegraphics[width=0.45\textwidth]{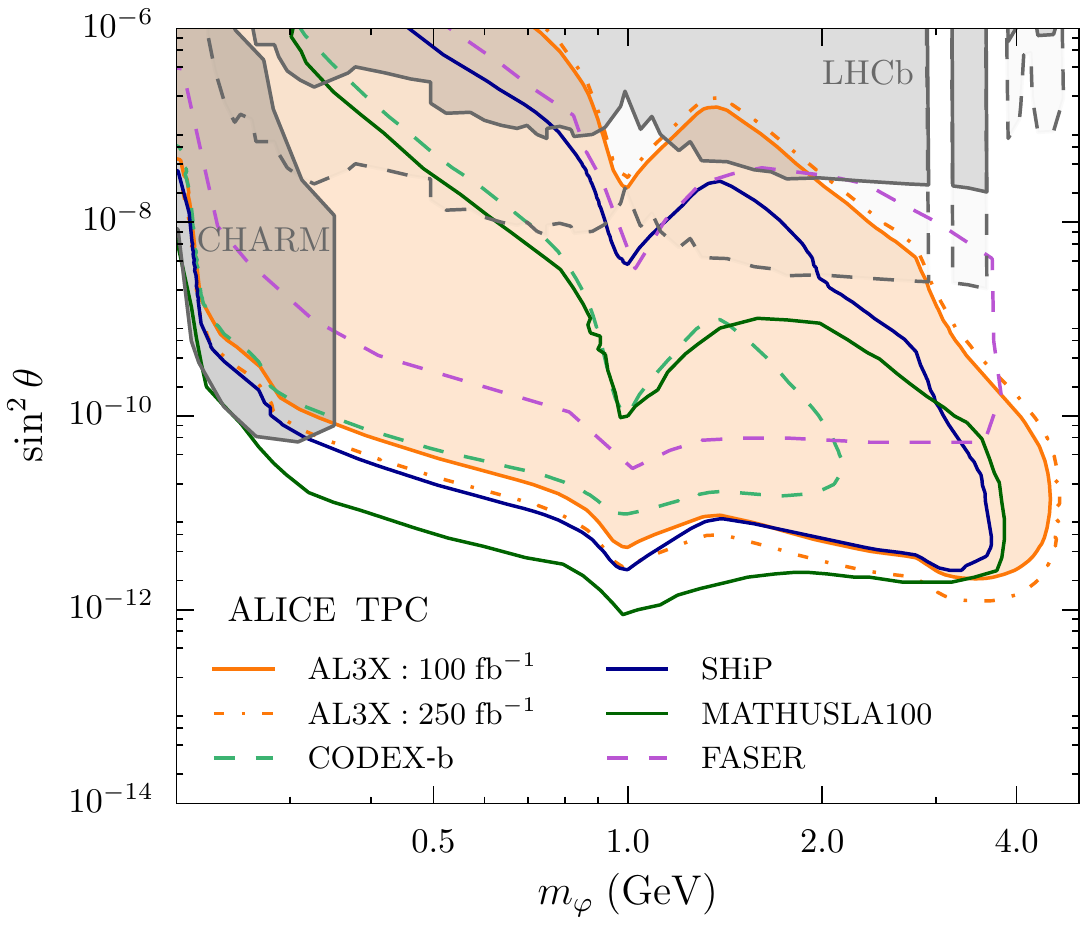}
\caption{AL3X reach for $B\rightarrow X_s\X$ in the $s^2_\theta$--$m_\X$ plane for a full detector (left) and with just the existing ALICE TPC (right). The remaining solid (dashed) curves indicate the various existing (projected) constraints from the various existing, past or proposed experiments, as described in the text. \label{fig:Bdecay}}
\end{figure*}

Fig.~\ref{fig:Bdecay} shows the reach in the Higgs mixing portal $s^2_\theta$--$m_\X$ parameter space, assuming 95\%CL exclusion and negligible expected backgrounds. The existing constraints are from CHARM~\cite{BERGSMA1985458} and LHCb~\cite{Aaij:2016qsm}; we also show the projected reach for LHCb, SHiP~\cite{Lanfranchi:2243034}, MATHUSLA~\cite{Evans:2017lvd}, CODEX-b~\cite{Gligorov:2017nwh} and FASER~\cite{Feng:2017vli} for comparison. The LHCb reach was estimated by rescaling the $B\rightarrow K (\X \rightarrow \mu\mu)$ limit~\cite{Aaij:2016qsm}, optimistically assuming zero background. When needed, the reach of other proposals was recasted to match the assumptions for the lifetime of $\X$ in Refs~\cite{Fradette:2017sdd,Bezrukov:2009yw}.

\subsection{Exotic $\pi^0$ and $\eta$ decays\label{sec:darkphoton}}

The final LLP production scenario we consider is via exotic decays of unflavored mesons. A popular benchmark for this type of process is a light $U(1)$ gauge boson that kinetically mixes with the SM photon through the operator $\epsilon /2\, F^{\mu\nu} F'_{\mu\nu}$. The $\pi^0$ and $\eta$ branching ratios are \cite{Batell:2009di}
\begin{align}
\text{Br}(\pi^0 \to \gamma A')= 2\epsilon^2 \left(1-\frac{m_{A'}^2}{m_{\pi^0}^2}\right)^3 \text{Br}(\pi \to \gamma\gamma)\\
\text{Br}(\eta \to \gamma A')= 2\epsilon^2 \left(1-\frac{m_{A'}^2}{m_{\eta}^2}\right)^3 \text{Br}(\eta \to \gamma\gamma).
\end{align}
The $\eta'\to \gamma A'$ and $\omega \to \pi^0 A'$ processes may be also considered, but we have verified that do not contribute substantially to the sensitivity. The $A'$ width is given by 
\begin{equation}
\Gamma =\frac{1}{3}\epsilon^2 \alpha m_{A'}\sqrt{1-\frac{4m_e^2}{m_{A'}^2}} \times \frac{1}{\text{Br}(A' \to e^+ e^-)}.
\end{equation}
where we take the branching ratio to electrons from Ref.~\cite{Buschmann:2015awa}. The lifetime of the dark photon is therefore only prolonged by the smallness of the mixing parameter ($\epsilon \ll1$). The long lifetime regime is therefore only accessible at the LHC because of the huge numbers of $\pi^0$ and $\eta$ mesons that are produced with a relatively high boost. 

Searches for a kinematically-mixed dark photon have been conducted for several decades, leading to a considerable list of existing constraints from a variety of probes. In the high $\epsilon$, short lifetime regime, the dominant constraints come from  high intensity lepton colliders and B-factories:
A1~\cite{Merkel:2014avp}, 
APEX~\cite{Abrahamyan:2011gv}, 
BaBar~\cite{Lees:2014xha}, 
KLOE \cite{Anastasi:2015qla,Anastasi:2016ktq,Babusci:2012cr,Babusci:2014sta,Babusci:2014sta}
LHCb \cite{Aaij:2017rft,Aaij:2017rft}. (Several of the existing constraints were computed with the \mbox{DarkCast} package \cite{Ilten:2018crw}). 
The low $\epsilon$, high $c\tau$, part of the dark photon parameter space has been probed by a range of beam dump and neutrino experiments: LSND~\cite{Athanassopoulos:1997er,Batell:2009di,Essig:2010gu}, 
CHARM~\cite{Bergsma:1985qz}, 
SLAC beam dumps \cite{Bjorken:1988as,Riordan:1987aw,Bross:1989mp}, 
KEK \cite{Konaka:1986cb}, 
NA48 \cite{Batley:2015lha}, 
NA64 \cite{Banerjee:2018vgk}, 
NOMAD \cite{Astier:2001ck},
 $\nu$CAL\cite{Blumlein:1990ay,Blumlein:1991xh} and 
ORSAY \cite{Davier:1989wz}.
 Finally, the very low $\epsilon$ regime is constrained by limits on the anomalous cooling of supernova SN1987a \cite{Chang:2016ntp}.

\begin{figure*}\centering
\includegraphics[width=0.45\textwidth]{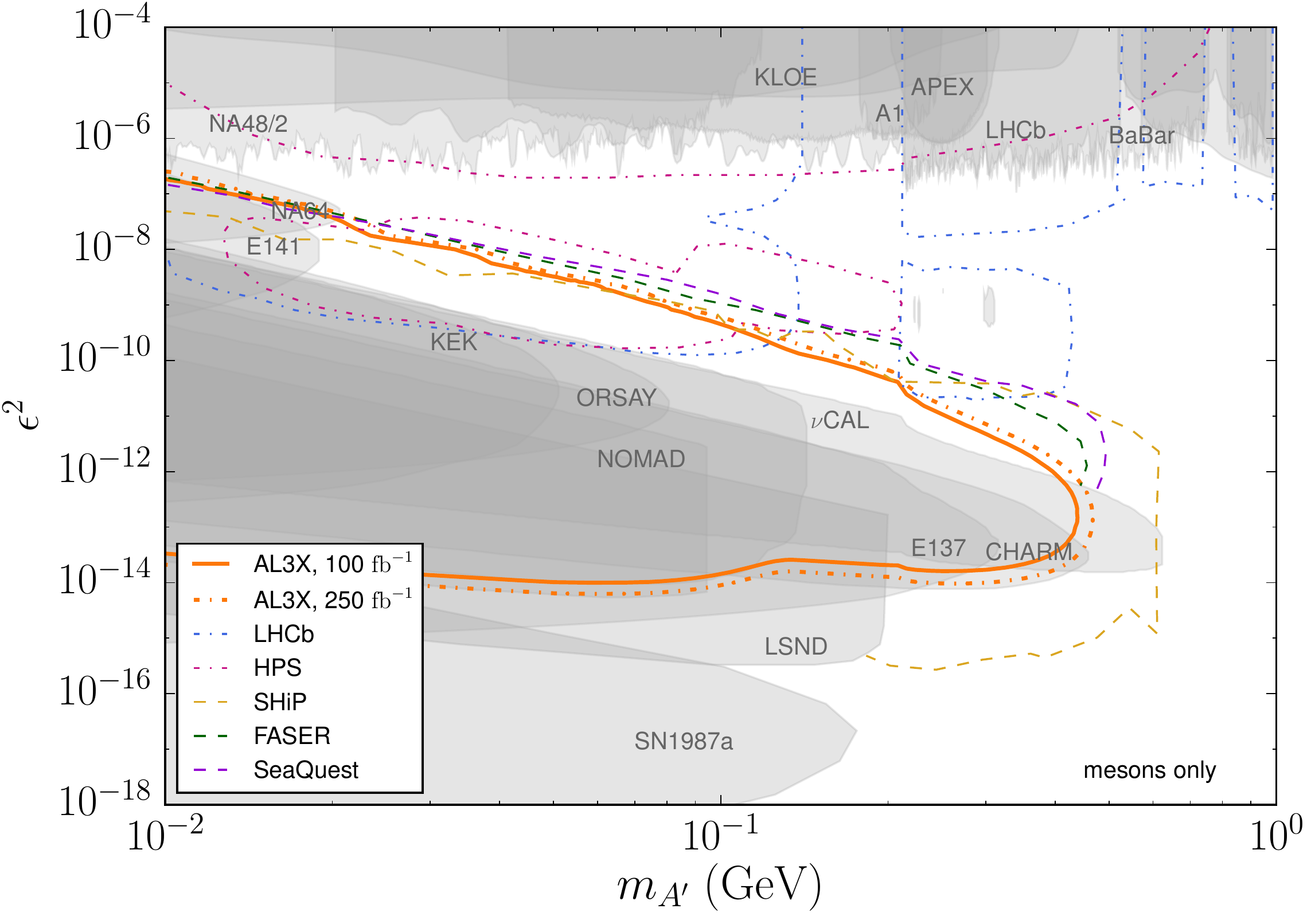}\hfill
\includegraphics[width=0.45\textwidth]{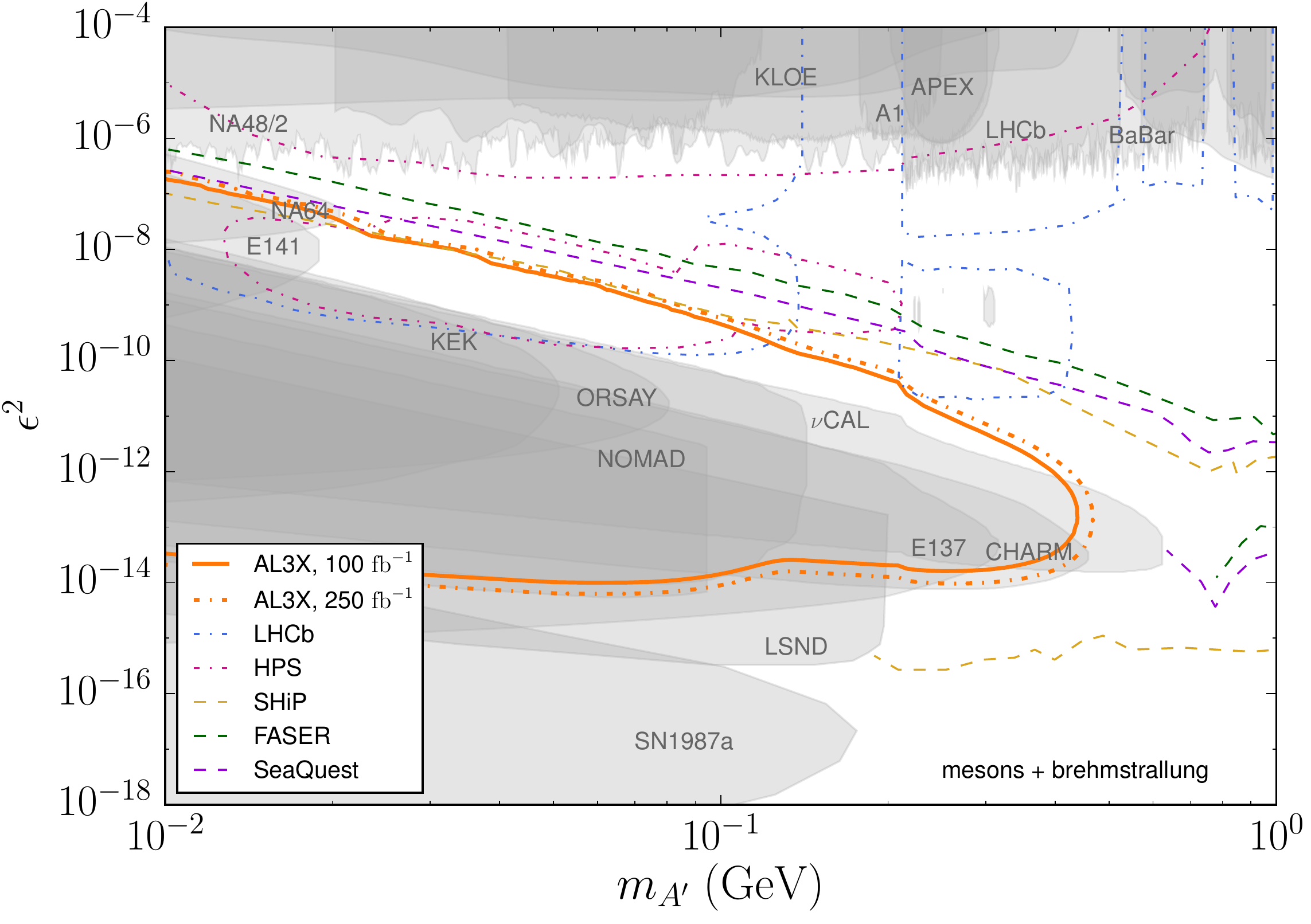}
\caption{
 \textbf{(left)} Existing constraints and projected reach for a kinetically mixed dark photon, including only meson decays.  \textbf{(right)} Same projections and constraints, but for SeaQuest, SHiP and FASER the bremsstrahlung has been included. 
 \label{fig:darkphoton}}
\end{figure*}

We compute the boost distribution of the $A'$ and the geometric acceptance of AL3X, using a minimum bias sample generated with \texttt{Pythia~8}, using the measured inelastic cross section of 68\,mb \cite{Aaboud:2016mmw}.  To model the tail of the $A'$ boost distribution, we also include several weighted dijet samples with increasingly stringent cuts on the parton level process. Specifically, we demand that at least one hard parton satisfies $\eta<4$ and $p_T>30$ GeV, as well as a lower bound on its energy, where the latter is varied over different samples. The \texttt{Pythia} level cross sections are corrected with a $\kappa$-factor of $1.1$, by comparing with the corresponding measurements \cite{Aad:2011fc}.  To compute the detector efficiency, we add the efficiencies obtained from each of these samples, weighted by the appropriate fiducial cross section. The resulting reach is shown in Fig.~\ref{fig:darkphoton}.  Also shown are the aforementioned existing constraints as well as the projected limits from planned or proposed experiments like SeaQuest \cite{Gardner:2015wea,Berlin:2018pwi},  LHCb \cite{Ilten:2015hya,Ilten:2016tkc}, SHiP \cite{CERN-SHiP-NOTE-2016-004}, HPS \cite{Moreno:2013mja} and FASER \cite{Feng:2017uoz}.

We find that for exotic $\pi^0$ and $\eta$ decays, AL3X can probe new parameter space in the high $\epsilon$, low $c\tau$, regime, but does not exceed the reach of dedicated forward experiments like FASER, SHiP and SeaQuest. This can be understood as follows: Though AL3X would likely have much less proton collisions than FASER, SeaQuest and SHiP, it would have the shortest effective baseline. In other words, since AL3X is somewhat forward from the IP, the tail of the kinetic energy distribution of the $A'$ extents as high as several TeV. This is higher than SeaQuest and SHiP, and somewhat lower than FASER, though FASER would be located \mbox{$\sim$~400 m} from the IP, compared to $\sim 5$ m for AL3X.

 For the kinetically mixed dark photon model in particular, FASER, SeaQuest and SHiP also have reach beyond the $\eta$ mass due to $A'$ production through bremsstrahlung of the protons\footnote{There is also an inelastic contribution from $A'$ plus jet(s) production, which is difficult to reliably quantify in the very forward regime (FASER, SeaQuest and SHiP). AL3X is located centrally enough that this process can be computed with standard tools, and we have verified that this contribution is negligible.}, which is not available to AL3X.  It should be mentioned that at AL3X a large flux of charged hadrons impinge on the shield, and it is possible this could result in a bremsstrahlung contribution as well. We save this calculation for future work. For reference, the FASER, SeaQuest and SHiP reach, including this additional contribution is shown in the right-hand panel of Fig.~\ref{fig:darkphoton}. 
 
We conclude that AL3X is competitive with the dark photon reach obtainable by HPS, LHCb, SeaQuest and FASER, though these either exist already in some form, or are substantially smaller in scale than AL3X. 
While AL3X is then likely not the most ideal configuration for the specific case of the kinetically mixed dark photon, these results nonetheless demonstrate that AL3X has non-negligible reach for low mass LLP portals.

\section{Discussion\label{sec:discussion}}
The potential strength of an experiment like AL3X lies in its versatility over a wide mass range of LLP portals: 
Thanks to its short baseline near an LHC interaction point, it can probe a range of LLP production modes all the way from exotic pion decays up to high mass portals like Higgs or top decays, or even heavier exotic states. 
We have seen that the AL3X concept can significantly increase the LLP reach over a broad range of portals, with respect to ATLAS, CMS and LHCb. 
The AL3X configuration is in essence a tracking detector behind a heavy shield, which can be thought of as analogous to a calorimeter that is solely absorber. 
While ATLAS, CMS and LHCb do have tracking stations (muon spectrometers) following their calorimeters, their hadronic calorimeters are much thinner than the shield in the AL3X design. 
This permits AL3X to search for much rarer signals in a very low background environment compared to ATLAS and CMS, and in this sense AL3X would be complementary to the existing 
(and proposed upgraded) multi-purpose detectors.

The AL3X configuration shown in Fig.~\ref{drawing} only makes use of collision debris for $\eta > 0.8$.  This leaves the entire region with $\eta < 0.8$ available for another experiment to co-exist. One such possibility could be a dedicated detector to study $\tau$-neutrinos, along the lines of what is proposed for SHiP \cite{Anelli:2015pba}. We leave the exploration of such a companion experiment to future studies. To avoid challenges of moving the IP, an alternative configuration for AL3X could be to keep IP2 fixed and build a cylindrical shield and detector around the interaction point within the L3 magnet itself.  Such a configuration would have smaller, lower energy backgrounds as it is more central, and its signal acceptance would be roughly comparable to configuration in Fig.~\ref{drawing}.  However, the restricted radial size inside the L3 magnet -- at most, $5.9$\,m -- would require a tungsten-only shield to preserve enough space for a viable TPC detector, and the larger angular size would require a much larger, and heavier, amount of tungsten than in Fig.~\ref{drawing}. There are also significant engineering challenges with this concept due to the mechanical weakness of the L3 magnet: Both the shield and the detector would effectively need to be suspended inside the magnet.   

To fully establish the feasibility and cost estimate of an experiment like AL3X at IP2, a number of things should be studied further. First and foremost, a study of the required beam configuration is needed to establish the technical feasibility and cost of delivering more luminosity,  needed for virtually any new experiment looking for beyond the Standard Model physics at IP2. Next, a more detailed simulation of the shield configuration and the detector response is required, possibly by making use of some of the existing ALICE reconstruction software. Including the background rejection power of the detector itself, such a simulation will inform a more realistic design and size for the shield, compared to our conservative estimates in this analysis. Simultaneously, on the theory side, the reach of a number of additional benchmark models -- e.g. dark matter model(s), axion-like particles, and heavy neutral leptons -- can be evaluated~\cite{Evans:2018ip}.

\begin{acknowledgments}
We thank Anthony Barker, Asher Berlin, David Curtin, Nathaniel Craig, Jared Evans, Maurice Garcia-Sciveres, Andy Haas, Peter Jacobs, Spencer Klein, Zoltan Ligeti, Peilian Liu, Harikrishnan Ramani, Simone Pagan-Griso, Mateusz Ploskon, Scott Thomas, Jessie Shelton, David Stuart and Jose Zurita for useful discussions.  We are particularly grateful to Sergey Antipov, John Jowett, Gianluigi Arduini, and Burkhard Schmidt for helping us understand some of the possible accelerator related challenges, to Peter Jacobs for supplying us with details on the ALICE cavern and setup and to David Curtin, Maurice Garcia-Sciveres, Peter Jacobs and Spencer Klein for providing valuable comments on the manuscript. 
VVG acknowledges funding from the European Research Council (ERC) under the European Union's Horizon 2020 research and innovation programme under grant agreement No 724777 ``RECEPT''. 
SK and MP are supported in part by the LDRD program of LBNL under contract DE-AC02-05CH11231, and by the National Science Foundation (NSF) under grants No.~PHY-1002399 and PHY-1316783. SK also acknowledges support from DOE grant DE-SC0009988 and from the Kavli Institute for Theoretical Physics, supported in part by the National Science Foundation under Grant No.~NSF PHY-1748958, where part of this work was performed. 
The work of DR was supported in part by NSF grant PHY-1720252. DR also acknowledges support from the University of Cincinnati, and thanks the Aspen Center of Physics, supported by the NSF grant 
PHY-1607611, where parts of this work were completed.    
This research used resources of the National Energy Research Scientific Computing Center, which is supported by the Office of Science of the DoE under Contract No.~DE-AC02-05CH11231.

\end{acknowledgments}

\bibliographystyle{apsrev4-1}
\bibliography{al3x}

\appendix

\section{Particle Gun Grids}
\label{sec:grids}

\begin{sidewaysfigure}[p]
\centering
\vspace{7cm}
\includegraphics[width=\textwidth]{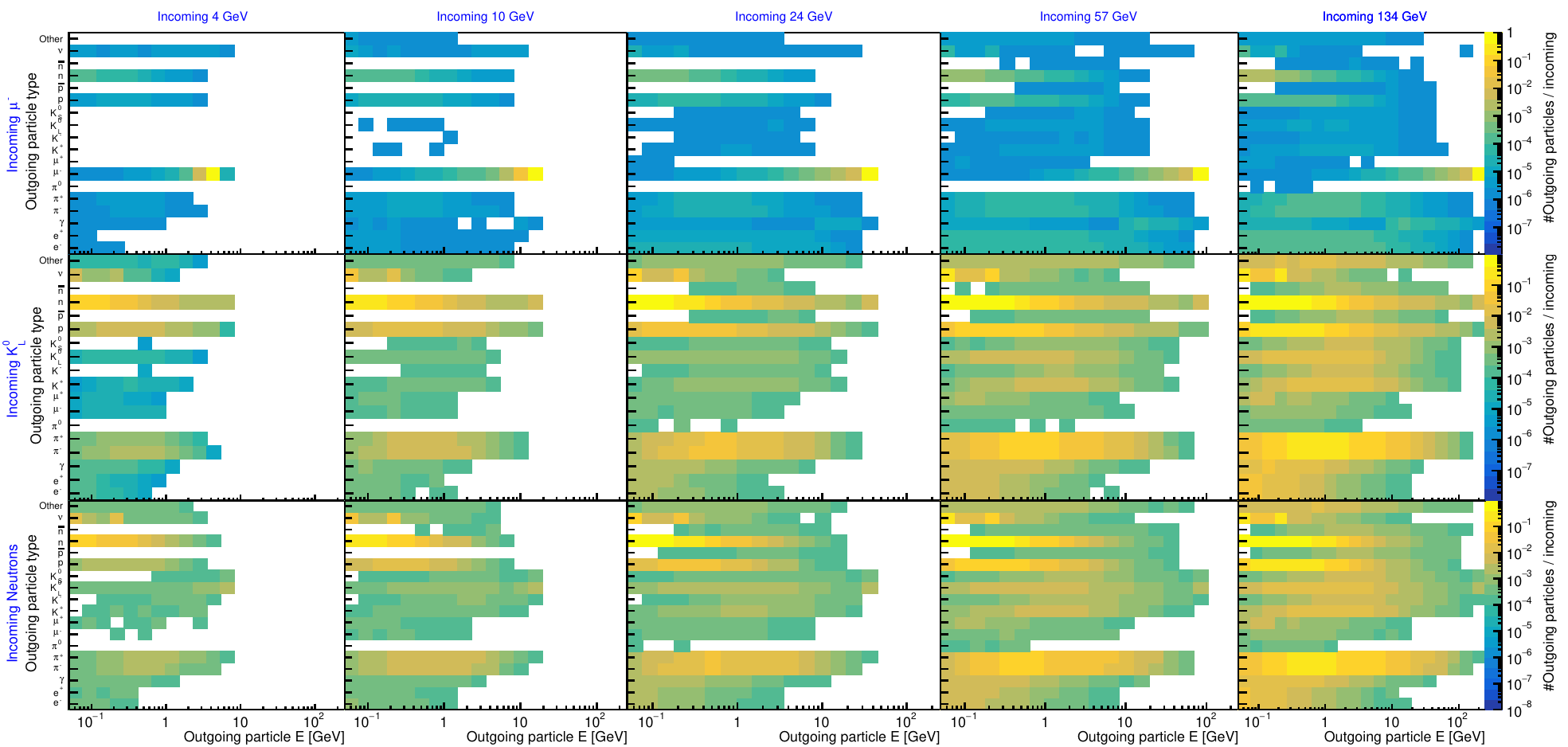}\
\caption{
The outgoing particle spectrum after $5\lambda$ of W for an incoming $\mu^-$ (top row), $K_L^0$ (middle row), and neutrons (bottom row) at various energies.  These are a selection of 15 grids from a total of $18\times 20$: 18 species, logarithmically spaced in energy from 50 MeV to 300 GeV in 20 bins.
 \label{fig:grids}}
 \end{sidewaysfigure}

Sample grids are shown in Fig.~\ref{fig:grids} for muons, $K_L^0$, and neutrons.  The full background estimation includes similar grids for each incoming particles species and energy, for a total of $18\times 20$ grids to account for all 18 incoming particle species and energies logarithmically spaced in incoming particle energy between 50 MeV and 300 GeV.  The transfer probability for $K_L$'s for various sample grid points was validated against the measured values in Ref.~\cite{PhysRevLett.42.9} using a Pb shield, while the muon propagation is consistent with CSDA estimates \cite{PDG:2016}.


\end{document}